\documentclass[aps,pre,onecolumn,superscriptaddress]{revtex4}
\usepackage[dvips]{graphics}
\usepackage{graphicx}
\usepackage{amsfonts}
\usepackage{amssymb}
\usepackage{amsmath}
\usepackage{subfigure}

\begin{document}

\title{
Matter-wave solitons with a periodic, piecewise-constant nonlinearity
}

\author{A.\ S.\ Rodrigues}
\affiliation{Departamento de F\'{\i}sica/CFP, Faculdade de Ci\^{e}ncias, Universidade do Porto, R. Campo Alegre,
687 - 4169-007 Porto, Portugal}

\author{P.\ G.\ Kevrekidis }
\affiliation{Department of Mathematics and Statistics, University of Massachusetts,
Amherst MA 01003-4515, USA}

\author{Mason A.\ Porter}
\affiliation{Oxford Centre for Industrial and Applied Mathematics, Mathematical Institute, University of Oxford, OX1 3LB, United Kingdom}

\author{D.\ J.\ Frantzeskakis}
\affiliation{Department of Physics, University of Athens, Panepistimiopolis, Zografos, Athens 157 84, Greece}

\author{P.\ Schmelcher}
\affiliation{Theoretische Chemie, Physikalisch-Chemisches Institut, Im Neuenheimer Feld 229,
Universit\"at Heidelberg, 69120 Heidelberg, Germany}
\affiliation{Physikalisches Institut, Universit\"at Heidelberg, Philosophenweg 12, 69120 Heidelberg, Germany}

\author{A.\ R.\ Bishop}
\affiliation{Theoretical Division and Center for 
Nonlinear Studies, Los Alamos National Laboratory, Los Alamos, New Mexico 87545, USA}

\begin{abstract}

Motivated by recent proposals of  ``collisionally inhomogeneous'' Bose-Einstein condensates (BECs), 
which have a spatially modulated scattering length, we study the existence and stability properties of bright and dark matter-wave solitons of a BEC characterized by a periodic, piecewise-constant scattering length.  We use a ``stitching'' approach to analytically approximate the pertinent solutions of the underlying nonlinear Schr\"odinger equation by matching the wavefunction and its derivatives at the interfaces of the nonlinearity coefficient.  To accurately quantify the stability of bright and dark solitons, we adapt general tools from the theory of perturbed Hamiltonian systems. We show that solitons can only exist at the centers of the constant regions of the piecewise-constant nonlinearity.  We find both stable and unstable configurations for bright solitons and show that all dark solitons are unstable, with different instability mechanisms that depend on the soliton location. We corroborate our analytical results with numerical computations.

\end{abstract}

\maketitle



\section{Introduction}

In the past few years, the study of solitary wave structures in atomic physics has received 
considerable attention, predominantly because of impressive advances in the field of Bose-Einstein condensation \cite{book1,book2}. These investigations have largely been motivated by 
the comprehensive mean-field description of atomic Bose-Einstein condensates (BECs) by the Gross-Pitaevskii (GP) equation, a variant of the nonlinear Schr{\"o}dinger (NLS) equation in which the nonlinearity, whose strength is proportional to the $s$-wave scattering length, is introduced by the interatomic interactions. This remarkable feature has allowed both theoretical investigations and experimental observations of bright  \cite{expb1,expb2,expb3}, dark \cite{dark1,dark2,dark3,dark4}, and gap \cite{gap} matter-wave solitons in BECs (see also the recent review \cite{ourbook}).  Such solitons have been studied in detail in the literature in the presence of various external potentials, including harmonic traps and periodic lattices.  This has yielded tremendous insights into a large variety of interesting phenomena, including Bloch oscillations, Landau-Zener tunneling, modulational (``dynamical") instabilities, gap excitations, and more (see the reviews \cite{ourbook,pgk,konotop,blochol,morsch} and references therein).  The types of matter-wave solitons that arise in a given situation depends on the nature of the interatomic interactions which, in turn, are characterized by the sign of the scattering length. Specifically, bright 
solitons arise in BECs with attractive interactions and negative scattering length, 
whereas dark and gap solitons emerge in the case of repulsive interactions and 
positive scattering length. Gap solitons also require the presence of an optical lattice or superlattice potential.  

During the past decade of BEC research, a number of tools have been developed to control and manipulate matter waves.  For example, they can be influenced, processed, and shaped using static (homogeneous and inhomogeneous) electric and magnetic fields \cite{Folman}, optical devices \cite{Grimm}, and near-field radio-frequency devices 
\cite{Lesanovsky}.  One can manipulate not only a BEC's external (trapping) potential but also the interactions among the atoms that are responsible for the nonlinear 
properties and dynamics of the matter waves. 
The interaction among ultracold atoms can be adjusted experimentally by employing
either magnetic \cite{Koehler,feshbachNa} or optical Feshbach resonances \cite{ofr} in a very broad range.  Additionally, so-called ``confinement induced resonances" \cite{olshanii1998a,guenter} allow one to vary the effective one-dimensional (1D) coupling constant of quasi-1D systems by adjusting the transversal confinement length.  One can consequently vary the external potential  
while independently and simultaneously changing the strength of the nonlinearity 
by tuning the interatomic interactions.

The manipulation of BECs using Feshbach resonances has propelled a significant number of investigations that have been subsequently refined substantially.  Experimental achievements include the formation of bright matter-wave solitons and soliton trains for $^{7}$Li \cite{expb1,expb2} and $^{85}$Rb \cite{expb3} atoms by tuning the interatomic interaction within a stable BEC from repulsive to attractive, the formation of molecular condensates \cite{molecule}, and the probing of the BEC-BCS crossover \cite{becbcs}.  Moreover, theoretical studies have predicted that a time-dependent modulation 
of the scattering length can be used to stabilize attractive two-dimensional BECs against collapse \cite{FRM1} or create robust matter-wave breathers in 1D BECs \cite{FRM2}.  We remark in passing that relevant contexts involving dependence of the nonlinearity coefficient on the evolution variable
have also been of interest in nonlinear optics in describing pulse propagation in an optical fiber in the presence of periodically varying dispersion and nonlinearity \cite{medv1,medv2} (see also the references therein).  In that context, a popular approach has been to look for an averaged (over the fast variation scale) equation with constant coefficients in which solitary-wave solutions could be sought.  This approach has also been useful in the context of BECs in describing the averaged properties of solitary waves in the presence of a periodic, time-dependent modulation of the scattering length \cite{dep1,dep2,dep3}.

In addition to the aforementioned studies involving temporal variations of the interaction strength, 
it has been recently found that atomic matter waves exhibit novel features under the influence of 
a spatially varying scattering length and, consequently, a spatially varying nonlinearity. 
The resulting so-called ``collisionally inhomogeneous'' environment 
provides a variety of interesting and previously unexplored 
dynamical phenomena and potential applications, including adiabatic compression of matter waves \cite{our1,fka}, 
atomic soliton emission and atom lasers \cite{vpg12}, 
enhancement of the transmittivity of matter waves through barriers \cite{our2,fka2}, 
dynamical trapping of matter-wave solitons \cite{our2}, 
stable condensates exhibiting both attractive and repulsive interatomic interactions \cite{chin}, and the
delocalization transition of matter waves~\cite{LocDeloc}. 
Particular inhomogeneous frameworks that have been investigated include linear \cite{our1,our2}, parabolic \cite{yiota}, random \cite{vpg14}, 
periodic \cite{vpg16,LocDeloc,BludKon}, and localized (step-like) \cite{vpg12,vpg17,vpg_new} spatial variations.  There have also been a number of detailed mathematical studies
\cite{key-2,key-4,vprl}.  In particular, Refs.~\cite{key-2,key-4} examined
the effects of a ``nonlinear lattice potential" (i.e., a spatially periodic nonlinearity 
coefficient) on the stability/instability of solitary waves and elucidated the interplay between drift and diffraction/blow-up instabilities.  More recently, the interplay of nonlinear and linear potentials has been 
examined in both continuum \cite{ckrtj} and discrete \cite{blud_pre} settings.

Motivated by the above studies, as well as the suggestion of a piecewise-constant inhomogeneous 
scattering length in Refs.~\cite{vpg12,vpg_new}, we investigate in the present work
matter-wave solitons in the framework of the NLS equation with a periodic, piecewise-constant nonlinearity coefficient.  Preparing such a setup experimentally is challenging but possible in principle. For example, it could be achieved by integrating optical and magnetic traps on an atom chip \cite{Folman}. The underlying idea is as follows: An on-chip fiber-optic configuration generates a microscopic dipole waveguide with a strong transversal confinement \cite{eriks}.
This waveguide is augmented by a periodic array of either permanent magnets \cite{shevchenko} or current-carrying wires \cite{Folman} that create alternating zones of approximately zero and constant magnetic fields in the longitudinal direction. Tuning the magnetic field (via, for example, the currents in the wires) such that the resulting field strength leads to an atomic scattering process close to a Feshbach resonance allows one to achieve a substantial variation of the atomic scattering behavior in the form of an alternating, approximately piecewise-constant scattering length.  Refs.~\cite{vpg12,vpg_new} considered a localized region with positive scattering length (the rest of the BEC was noninteracting, providing a substantial linear regime).  Here, however, we consider a scattering length given by a piecewise-constant function that switches periodically between either two positive or two negative values.  The importance of this particular fully nonlinear setting is that it allows us to analytically investigate the existence and stability of matter-wave solitons in each region (where the nonlinearity coefficient is constant) and to subsequently ``stitch'' the constructed (local) solutions at the jump points of the piecewise-constant nonlinearity function.  Our analysis is similar in spirit to the approach used in a situation with both linear and nonlinear, piecewise-constant periodicity in Ref.~\cite{kominis}. We employ techniques from the general theory of perturbed Hamiltonian systems, starting from the unperturbed limit of the completely integrable 1D NLS equation, in order to analytically address the stability of both bright \cite{todd_bright} and dark \cite{PelKev07} matter-wave solitons.  In order to quantify the accuracy of our theoretical predictions, we subsequently compare these findings with direct numerical computations.

Our presentation is organized as follows. In Section \ref{setup}, we present our setup and the pertinent GP model.  We study bright and dark matter-wave solitons, respectively, in Sections \ref{bright} and \ref{dark}.  Finally, in Section \ref{conc}, we briefly summarize our findings and present our conclusions.

\section{Setup and model} \label{setup}

We consider a ``cigar-shaped" condensate that is elongated along the $z$ direction and strongly confined in the transverse ($x$ and $y$) directions . Assuming that this condensate is collisionally inhomogeneous, characterized by a spatially varying scattering length $a(z)$, we follow the standard approach \cite{gpe1d} of averaging the three-dimensional GP equation in the transverse plane.  We thereby arrive at the following 1D GP equation for the longitudinal wavefunction $\psi(z,t)$: 
\begin{equation}
	i \hbar \partial_{t}\psi(z,t)=\left[-\frac{\hbar^2}{2m} \partial_{z}^{2} 
	+ 2\hbar\omega_{\perp}a(z) \left|\psi(z,t)\right|^{2}\right]\psi(z,t)\,,
	\label{eq:GP0}
\end{equation}
where $m$ is the atomic mass and $\omega_{\perp}$ is the transverse confinenent frequency.

Measuring time and space in units of $\omega_{\perp}^{-1}$ and transverse oscillator length
$l_{\perp} \equiv \sqrt{\hbar/m \omega_{\perp}}$, respectively,
we reduce the GP in Eq.~(\ref{eq:GP0}) to the dimensionless form, 
\begin{equation}
	i\partial_{t}\phi(z,t)=\left[-\frac{1}{2}\partial_{z}^{2} + g(z)\left|\phi(z,t)\right|^{2}\right]\phi(z,t)\,,
	\label{eq:GP1}
\end{equation}
where $\phi(z)= \sqrt{2|a_{0}|} \psi(z)$ and $g(z)=a(z)/|a_0|$ is negative (positive) for attractive (repulsive) 
interatomic interactions.  In the above expressions, $a_0$ denotes the value of the scattering length 
of the corresponding collisionally homogeneous system, for which $g(z) = g_{0}=$ const. 
Note that Eq.~(\ref{eq:GP1}) has the form of a 1D NLS equation.  It possesses two integrals 
of motion: the normalized number of atoms $N=\int_{-\infty}^{+\infty}\left|\phi\right|^{2}dz$ and the Hamiltonian.
The former is connected to the unnormalized number of atoms $\cal{N}$ through the equation $\cal{N}$ $= (a_{\perp}/2|a_{0}|)N$. 

We take the nonlinearity coefficient $g(z)$ in Eq.~(\ref{eq:GP1}) to be a piecewise-constant function of space:
\begin{align}
	g(z) = \left\{ \begin{array}{cc}
	g_{0}\,, & {\rm if} \:0<{\rm mod}(z,L)<L_{1} \\
	g_{1}\,, & {\rm if} \: L_{1}<{\rm mod}(z,L)<L \end{array}\right.\,. \label{gofz}
\end{align}
Note that the nonlinearity coefficient in Eq.~(\ref{gofz}) may also be expressed as $g(z)=g_{0}+\Delta g(z)$, where
\begin{align}
	\Delta g(z)=\Delta {g_0}\sum_{n=-\infty}^{n=+\infty}\left\{ \Theta\left(z-\left[nL+L_{1}\right]\right)-\Theta\left(z-[n+1]L\right)\right\}\,, \label{gofz2}
\end{align}
where $\Delta {g_{0}}=g_{1}-g_{0}$ and $\Theta$ is the Heaviside step function. 

In our numerical investigations (see the discussion below), we adopt values of the parameters that are typical for experimental setups.  In particular, we consider a quasi-1D trap with a transverse confining frequency $\omega_{\perp}=2\pi \times 1400$ Hz, which fixes the temporal unit to $0.1$ ms. 
We assume that the dimensionless chemical potential $\mu$ is of order $O(1)$. 
For an (attractive) $^{7}$Li condensate, this choice yields a spatial unit of $1 \mu$m 
and a number of atoms of $N \approx 1100$.  For the (repulsive) $^{87}$Rb (respectively, $^{23}$Na) condensate, this yields a spatial unit of $0.3\mu$m ($2.2\mu$m) and a number of atoms of $N \approx 1200$ ($16000$).  Moreover, we assume that the nonlinearity function $g(z)$ is characterized by changes $\Delta g$ of the order of $50\%$ of the typical values of the scattering lengths, and that the normalized periodicity $L$ is of the order $O(1)$ (i.e., it is about a few microns).  This choice is consistent with typical Feshbach resonances that occur in Li, Rb, and Na condensates: Changing the magnetic field by a factor of two on a micron scale in the neighborhood of a 
Feshbach resonance requires magnetic field gradients in the range $1$ kG/cm to $100$ kG/cm, 
which are of moderate size for atom chips.  The lower limit 
is also compatible with the gradients used in macroscopic trap setups.

Below, we will quantify the existence and stability properties of bright and dark 
matter-wave solitons (for negative and positive nonlinearity coefficients $g(z)$, respectively) as a function of the inhomogeneity strength $\Delta {g_0}$.

\section{Bright matter-wave solitons} \label{bright}

\subsection{Perturbation analysis} \label{bright_perturb}

Let us first consider an attractive BEC characterized by a negative nonlinearity coefficient in Eq.~(\ref{eq:GP1}), so that $g_{0}<0$ and $|\Delta g(z)|<|g_0|$.  In the special case of a 
homogeneous nonlinearity, scaled such that its coefficient is $g_{0}=-1$, 
Eq.~(\ref{eq:GP1}) possesses a bright matter-wave soliton solution of  the form
\begin{equation}
	\phi_{bs}(z,t)=\eta\ \mathrm{sech}[\eta(z-z_{0})]\exp(i\mu t)\,,\label{eq:BS}
\end{equation}
where $\eta$ is the soliton's amplitude and inverse width, 
$z_{0}$ is the position of its center,  
and $\mu=-(1/2)\eta^{2}$ is its effective chemical potential.  We then consider the spatial inhomogeneity as a perturbation of strength $\epsilon \equiv \Delta {g_0}$. 
When $\epsilon \neq 0$, the integrability of the 
original NLS in Eq.~(\ref{eq:GP1}) is lost, but the system is still Hamiltonian. 
Its Hamiltonian functional is given by $H=H_{0}+\epsilon H_{1}$, where the unperturbed ($H_0$) and perturbed ($H_1$) parts of the Hamiltonian are given by
\begin{align}
	H_{0} &= \int_{-\infty}^{+\infty}\frac{1}{2}\left(\left|\partial_{z}\phi\right|^{2}-\left|\phi\right|^{4}\right)dz\,, 
\label{uh} \\ 
	H_{1} &= \int_{-\infty}^{+\infty}\frac{1}{2}\frac{\Delta g(z)}{\Delta {g_{0}}}\left|\phi\right|^{4}dz\,. 
\label{ph}
\end{align}
In accord with the general perturbation theory for Hamiltonian systems developed in Ref.~\cite{todd_bright}, the condition for the existence of the solution (\ref{eq:BS}) under the aforementioned perturbation is that it remains an extremum of the perturbation Hamiltonian $H_1$. 
In the present case,  it is easy to see by considering the dependence of $H_1$ on $z_0$ that this condition is satisfied provided $z_0$ is either at the center of a region with nonlinearity coefficient $g_0$ or (by symmetry) at the center of a region with nonlinearity coefficient $g_1$. Therefore, we can identify stationary solitary wave solutions centered at these points in the presence of the periodic, piecewise-constant nonlinearity.

Let's consider the stability of these bright matter-wave solitons. As is well-known, the unperturbed NLS equation is invariant under both translation and phase/gauge transformations.  If the perturbation is sufficiently small, instability can arise in the case of attractive nonlinearity only via the perturbation-induced breaking of one of these symmetries.  (Note that the continuous spectrum is bounded away from the origin and for small $\epsilon$ cannot give rise to an instability \cite{todd_bright}.)  Furthermore, it is clear that in the presence of a spatially-dependent nonlinearity, the U$(1)$ symmetry  (i.e., the phase
invariance) of the equation is preserved, so the stability ultimately should depend on the location of the translational eigenvalue.  The 
eigenvalues $\lambda$ (and the corresponding eigenfrequencies $\omega$, which satisfy $\omega^2=-\lambda^2$) can be obtained within the framework of the perturbation analysis of Hamiltonian systems in Ref.~\cite{todd_bright} (see also Ref.~\cite{yiota}) according to the equation 
\begin{equation}
	{\rm det}\left(\epsilon\mathbf{M}-\omega^{2}D\right)=0\,, \label{eq:BS_eig}
\end{equation}
where the matrices $\mathbf{M}$ and $\mathbf{D}$ are given by
\begin{align}
	\mathbf{M} &= \left(\begin{array}{cc}
\frac{\partial}{\partial z_{0}}\left\langle \frac{\delta H_{1}}{\delta\phi^{*}},\frac{\partial\phi_{bs}}{\partial z_{0}}\right\rangle  & 0\\
0 & 0\end{array}\right)\,, 
\label{mm} \\
	\mathbf{D} &= \left(\begin{array}{cc}
\left\langle \frac{\partial\phi_{bs}}{\partial z},-z\phi_{bs}\right\rangle  & 0\\
0 & -\left\langle \phi_{bs},\frac{\partial\phi_{bs}}{\partial\eta}\right\rangle \end{array}\right)\,. 
\end{align}
In the above equations, the star $*$ denotes complex conjugation, $\langle \,, \rangle$ denotes the inner product, and $\delta H_{1}/\delta \phi^{\star}$ is the functional (Fr\'{e}chet) derivative.  

The nonzero elements of the matrices $\mathbf{M}$ and $\mathbf{D}$ (i.e., $m_{11}$, $d_{11}$, and $d_{22}$) can then be calculated directly to obtain
\begin{align}  \label{summ}
	m_{11} =  &-\eta^{5}\sum_{n=-\infty}^{n=+\infty}\left\{ \sinh\left[\eta\left([n+1]L-z_{0}\right)\right]\mathrm{sech}^{5}\left[\eta\left([n+1]L-z_{0}\right)\right]\right.\\
 &  \left.-\sinh\left[\eta\left(L_{1}+nL-z_{0}\right)\right]\mathrm{sech}^{5}\left[\eta\left(L_{1}+nL-z_{0}\right)\right]\right\}\,, 
\end{align}
where $d_{11}=\eta$ and $d_{22}=-1/\eta$.  Equation (\ref{eq:BS_eig}) allows one to calculate the translational eigenfrequencies to leading order in $\epsilon$, yielding
\begin{align}
	\epsilon m_{11}-\omega^{2}\eta=0\,. \label{freq}
\end{align}
The sum in Eq.~(\ref{summ}) can then be calculated numerically. 
Note that this sum converges rapidly because its component functions decay exponentially, and  
only three terms ($n=-1,0, \:\mathrm{and}\:+1$) are needed to obtain a value that is accurate up to ten digits.  Below we will directly 
compare the analytical prediction provided by Eq.~(\ref{freq}) 
to the numerical results for different values of $\Delta {g_0}$ (and of the 
soliton's chemical potential $\mu$) obtained by computations.

\subsection{``Stitching'' of bright soliton solutions}

We can use specific features of this particular system to go beyond the above theory, which is valid for general perturbed Hamiltonian systems, and acquire a quantitative handle on the 
perturbed soliton profile. In particular, we use the fact that one can explicitly construct the solution of the steady state equation for each of the regions in which the nonlinear coefficient is constant. For sufficiently small $\epsilon$, we expect the solutions to deviate little from the solution $\phi_{bs}$. We then construct an analytical approximation to the perturbed solution by ``stitching'' together the functions that would be solutions to the respective homogeneous NLS equations, with the nonlinearity coefficients taking different constant values in different spatial regions.  To do this, we require that the wavefunction $\phi$ and its first spatial derivative $\partial_{z}\phi$ be continuous across the boundaries between regions of different $g$. 

Let us consider, in particular, the bright soliton solution of Eq.~(\ref{eq:GP1}) centered at $z_{m}$ with amplitude $\eta_{m}$ in a region with nonlinearity coefficient $g_{m}$: 
\begin{equation}
	\phi_{m}(z)=\frac{\eta_{m}}{\sqrt{g_{m}}}\mathrm{sech}[\eta_{m}(z-z_{m})]\,. \label{bszm}
\end{equation}
Starting with a known value of, say, $\eta_{m}$ and $z_{m}$, we then need to calculate all other $\eta_{j}$ and $z_{j}$ ($j\neq m$) as a function of the known ones.  Applying the aforementioned continuity conditions, we obtain
\begin{align}
	\frac{\eta_{m}}{\sqrt{g_{m}}}\mathrm{sech}[\eta_{m}(Z_{m}-z_{m})] &= \frac{\eta_{m+1}}{\sqrt{g_{m+1}}}\mathrm{sech}[\eta_{m+1}(Z_{m}-z_{m+1})]\,, 
\label{matching1} \\
-\frac{\eta_{m}^{2}}{\sqrt{g_{m}}}\mathrm{sech}[\eta_{m}(Z_{m}-z_{m})]\tanh[\eta_{m}(Z_{m}-z_{m})] & = & -\frac{\eta_{m+1}^2}{\sqrt{g_{m+1}}}\mathrm{sech}[\eta_{m+1}(Z_{m}-z_{m+1})]
\nonumber \\
&\times \tanh[\eta_{m+1}(Z_{m}-z_{m+1})]\,,
\label{matching2}
\end{align}
where $Z_{m}$ is the coordinate of the interface between regions $m$ and $m+1$.  Using Eq.~(\ref{matching1}), it is then possible to simplify Eq.~(\ref{matching2}) to get
\begin{align*}
	\eta_{m+1}(Z_{m}-z_{m+1}) =  {\rm arctanh}\left\{ \frac{\eta_{m}}{\eta_{m+1}}\tanh\left[\eta_{m}(Z_{m}-z_{m})\right]\right\}\,,
\end{align*}
from which we obtain $z_{m+1}$ once $\eta_{m+1}$ is known.  We can then proceed iteratively to obtain $(z_{m+2},\eta_{m+2})$, etc.  Note, however, that this method is not exact, 
as can be inferred by close inspection.  In particular, 
Eqs.~(\ref{matching1})-(\ref{matching2}) possess two unknowns, $z_{m+1}$ and $\eta_{m+1}$, provided $z_{m}$ and $\eta_{m}$ are given. However, in principle, the quantity $\eta_{m+1}$ is determined by the coefficient $g_1$ of the piecewise-constant nonlinearity (and the chemical potential $\mu$). 
Hence, the system is, in fact, overdetermined. Nevertheless, as long as $\mu$ is not 
too small (so that the width of the soliton is smaller than or comparable to
the size of a step in the periodic nonlinearity (see Fig.~\ref{fig1}), the above stitching procedure is fairly accurate, as will be discussed below when we present the results of our numerical simulations.


\subsection{Numerical Results}

\begin{figure}
\includegraphics[width=100mm,keepaspectratio]{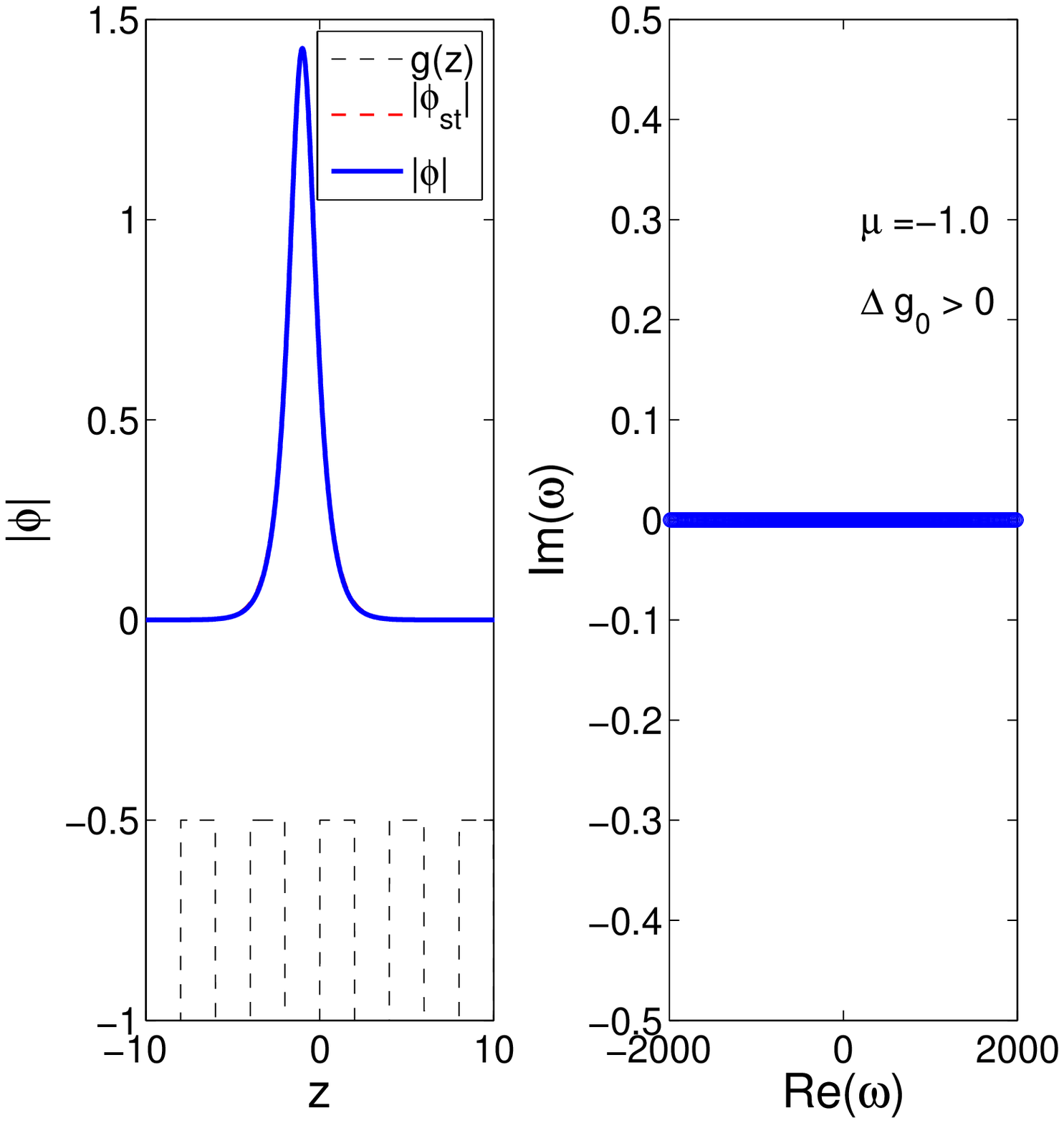}
\includegraphics[width=100mm,keepaspectratio]{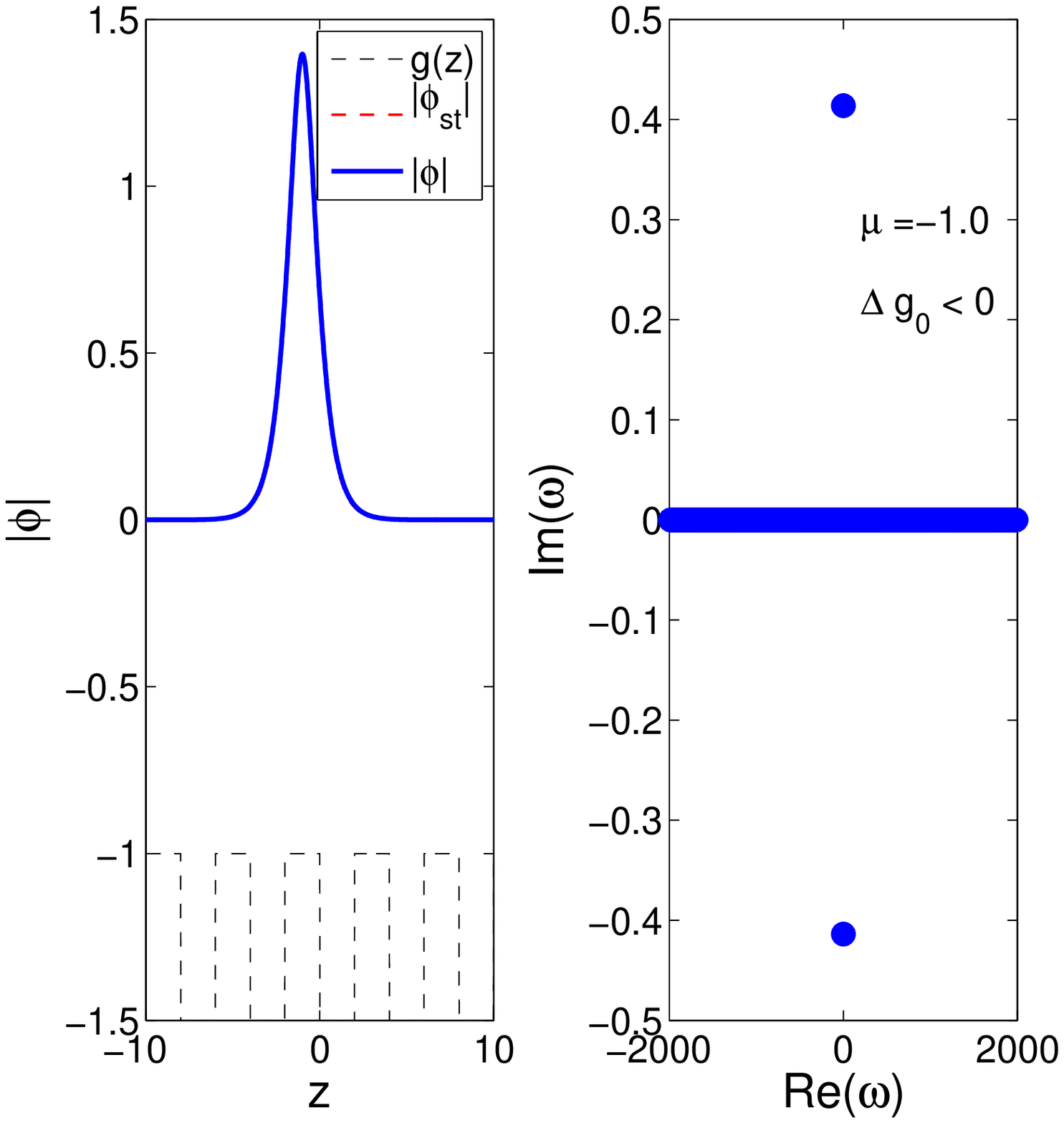}
\caption{(Color online) Stitched bright-soliton solutions versus fully numerical bright-soliton solutions of the NLS equation.  The top left panel shows the wave profile and the ``stitched'' solution for the same parameter values ($\mu=-1.0$ and $\Delta g>0$).  The top right panel gives the corresponding spectral plane $({\rm Re}(\omega),{\rm Im}(\omega))$ of the eigenfrequencies 
$\omega={\rm Re}(\omega)+ i {\rm Im}(\omega)$.  In the bottom panels, we show the same plots for a case with $\mu=-1$ and $\Delta g<0$.  The dashed curves in the left panels show the spatially dependent nonlinearity coefficients.  The bright soliton depicted in the top panels is stable, whereas that in the bottom panels is unstable. Notice that for these parameters the stitched solutions is
practically indistinguishable from the numerically exact one.
}
\label{fig1}
\end{figure}

\begin{figure}
\includegraphics[width=100mm,keepaspectratio]{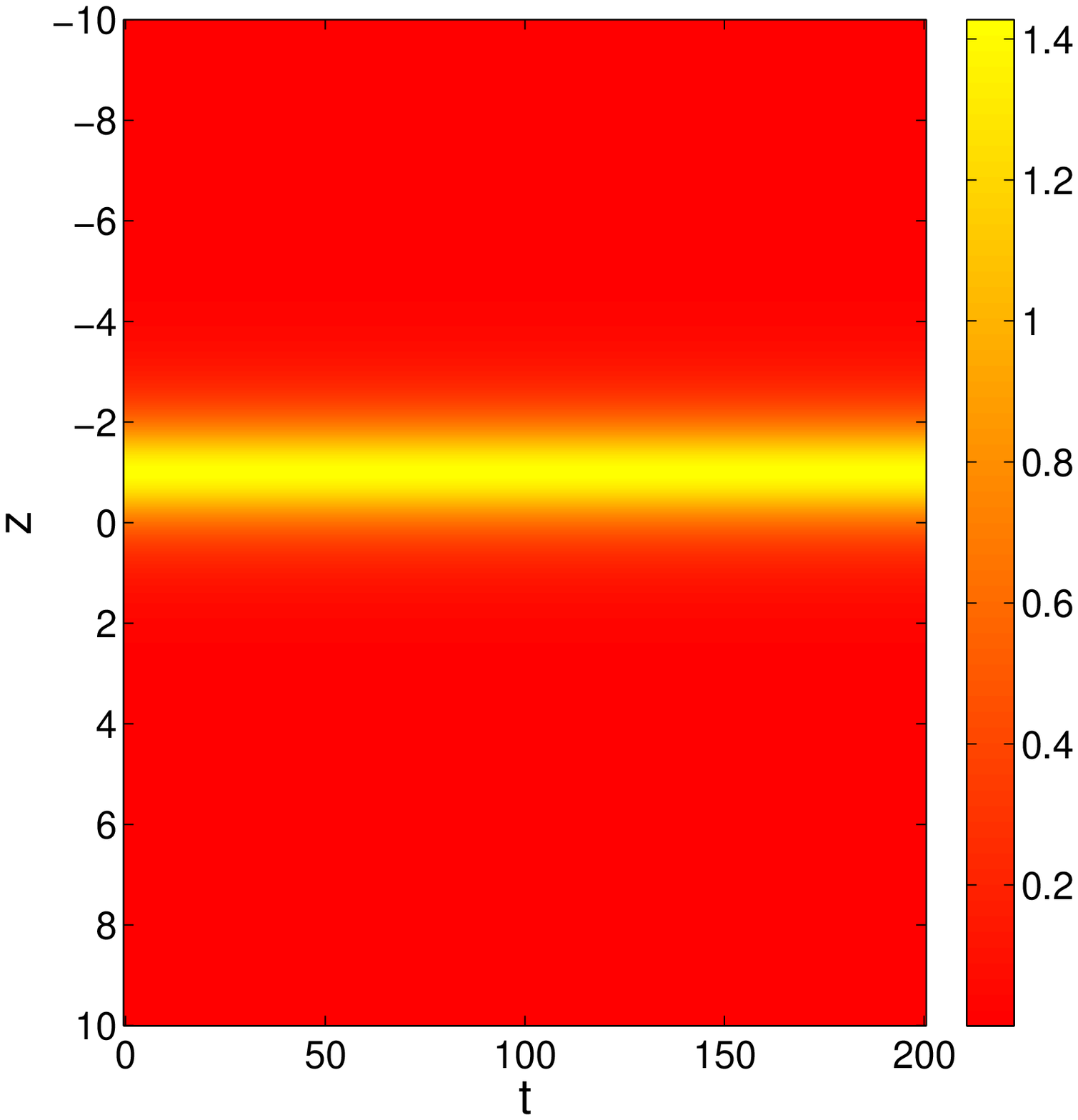}
\includegraphics[width=100mm,keepaspectratio]{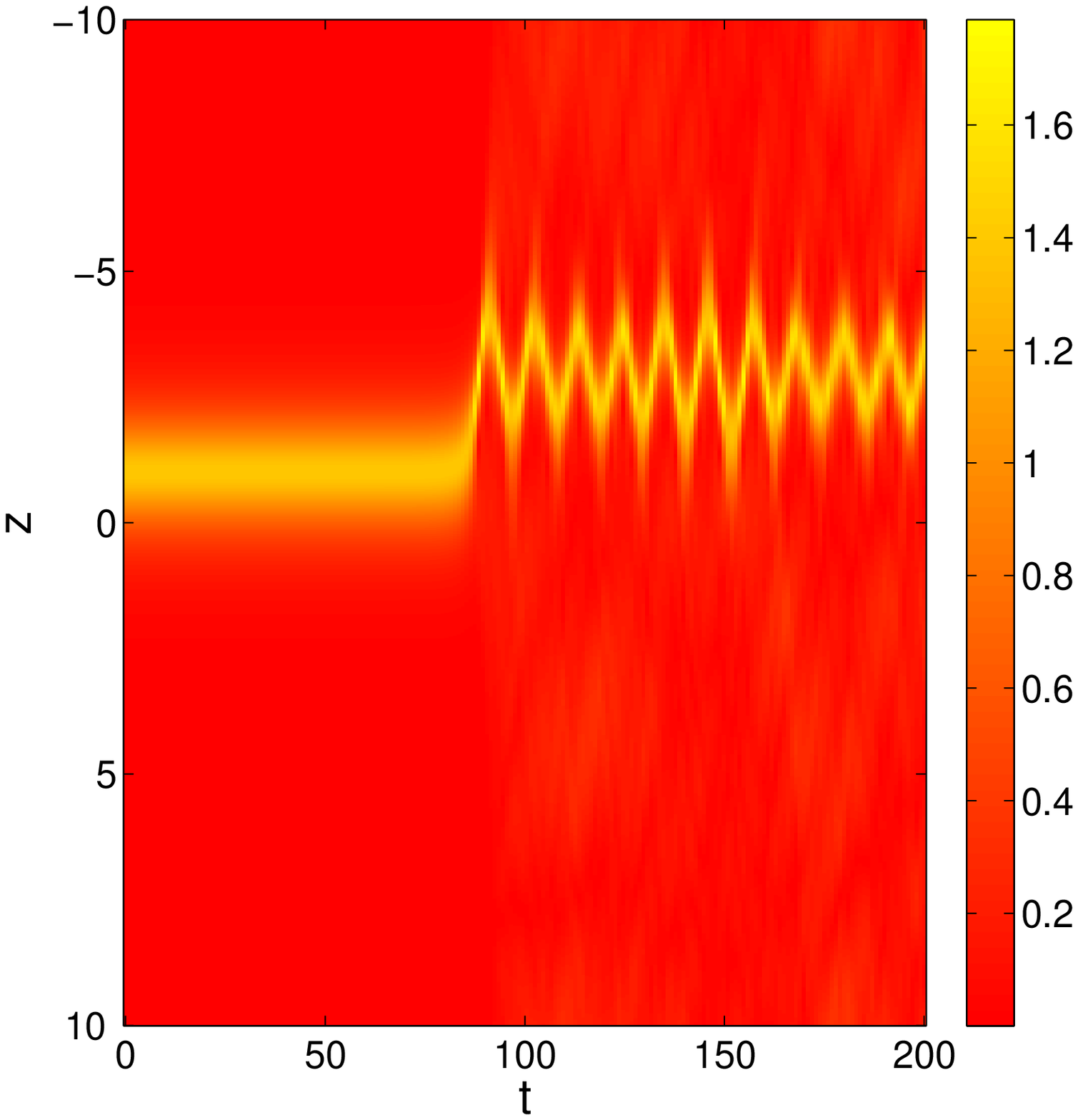}
\caption{(Color online) Time evolution of the bright matter-wave soliton for the two cases shown in Fig.~\ref{fig1}.  The top panel shows the contour plot of $|\phi(z,t)|$ for parameter values $\mu=-1.0$ and $\Delta g_0=+0.5$, whereas the bottom one shows the case of $\mu=-1.0$ and $\Delta g_0=-0.5$.  The bright soliton depicted in the top panel remains unchanged during the evolution, whereas that 
in the bottom panel drifts after $t \gtrsim 80$ and oscillates around a stable equilibrium
point (i.e., around a maximum of $|g(z)|$).
}
\label{figBdyn}
\end{figure}

We begin the description of our numerical findings by illustrating the solution obtained by numerically solving the standing wave problem corresponding to Eq.~(\ref{eq:GP1}) via a fixed point (Newton-Raphson) iteration.  In Fig.~\ref{fig1}, we show an example of both a stable bright matter-wave soliton solution (top left) and an unstable one (bottom left).  We also present  (in the right panels of the figure) the corresponding spectral planes $({\rm Re}(\omega),{\rm Im}(\omega))$ of the eigenfrequencies 
$\omega={\rm Re}(\omega)+ i {\rm Im}(\omega)$ that result from a Bogoliubov-de Gennes analysis 
of the NLS equation around the soliton solution. The absence of eigenfrequencies with nonzero imaginary part indicates stability, whereas the presence of imaginary or complex eigenfrequencies is a signature of linear instability.  The stability features observed are reminiscent of the ones found in earlier works, such as Refs.~\cite{key-2,key-4,ckrtj}.  Namely, the solution is stable when centered at a maximum of the magnitude (in absolute value) of the piecewise-constant, periodic nonlinearity, but it is linearly unstable when centered at a minimum.  In Fig.~\ref{fig1}, we also show the solutions that we obtained semi-analytically via the ``stitching'' procedure illustrated above.  Observe the excellent 
agreement between those solutions and those computed using fully numerical simulations. In fact, the two solutions are virtually indistinguishable in this case and, more generally, in all situations with sufficiently large $|\mu|$.

\begin{figure}
\includegraphics[width=100mm]{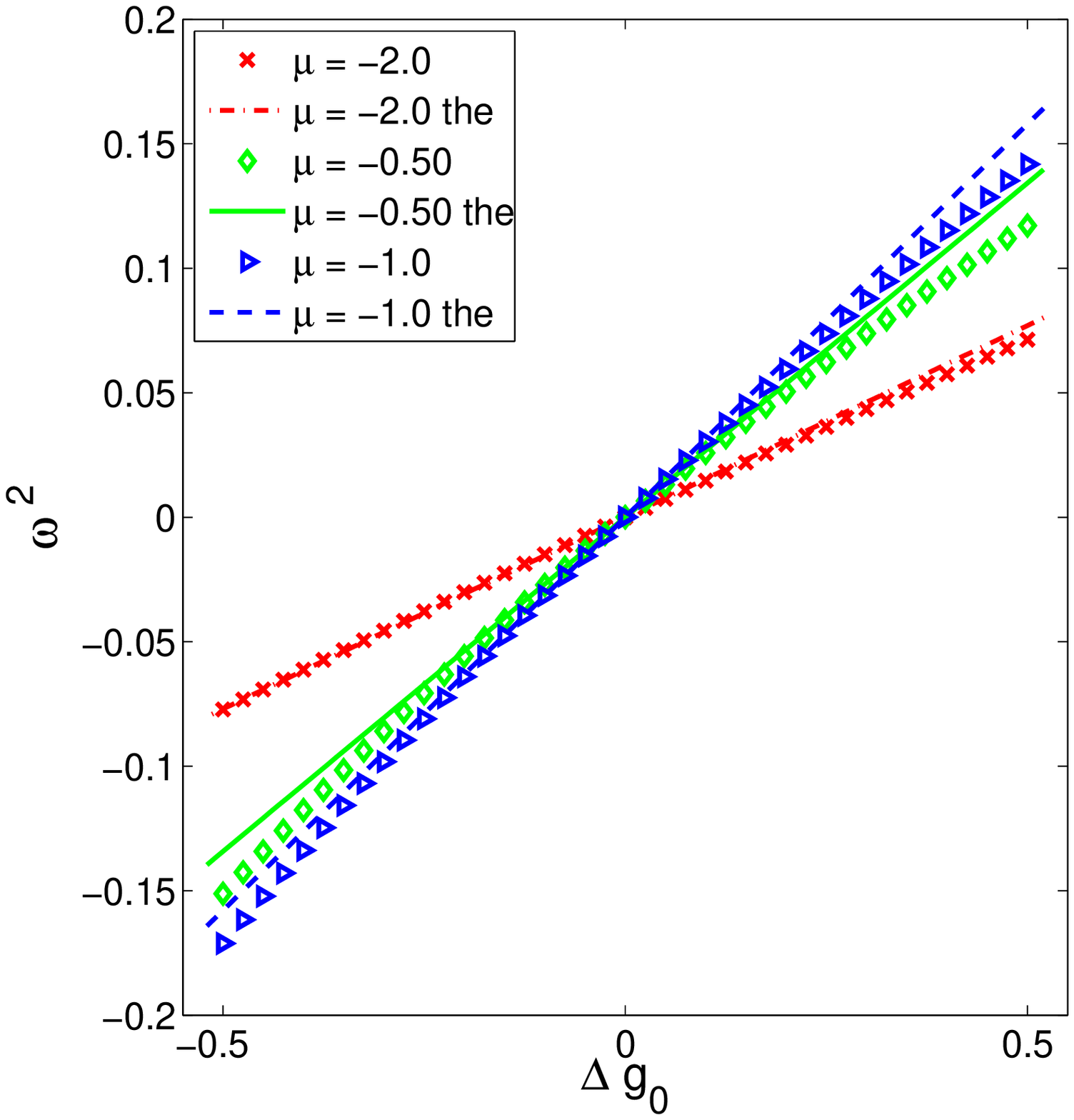}
\caption{(Color online) Numerical results (points) and theoretical predictions resulting from the perturbation analysis (curves, which are labeled ``the'' in the inset) for the translational squared eigenfrequency $\omega^2$ as a function of perturbation strength $\Delta {g_0}$ for different values of the effective chemical potential $\mu$ of the bright matter-wave soliton.
}
\label{fig2}
\end{figure}

Figure~\ref{figBdyn} depicts the evolution of bright matter-wave solitons with initial configurations
of the forms presented in Fig.~\ref{fig1}.  In particular, we present the stable case (which has parameter values $\mu=-1.0$ and $\Delta g_0=+0.5$) in the top panel.  Observe that the bright soliton remains unchanged during the evolution, as predicted by the linear stability analysis. On the other hand, we show the unstable case (with parameter values $\mu=-1.0$ and $\Delta g_0=-0.5$) in the bottom panel.  Observe that for $t \gtrsim 80$, the soliton drifts~\cite{key-2,key-4,ckrtj} and starts oscillating around a stable fixed point (the center of a neighboring maximum of the magnitude of the periodic
potential) as a result of the existence of the imaginary eigenfrequency.

In Fig.~\ref{fig2}, we show a conclusive diagram that summarizes our results for the linear stability of the bright-soliton solutions we constructed.  The plot shows the relevant translational (squared) eigenfrequency as a function of the strength of the perturbation $\Delta {g_0}$ for different values of the frequency $\mu$. It is clear that the corresponding squared eigenfrequency is essentially linear in $\Delta {g_0}$, giving rise to instability for $\Delta {g_0}<0$ and stability for $\Delta {g_0}>0$.  In all cases, we observe very good agreement between the theoretical prediction and the numerical results, especially when $\Delta {g_0}$ is sufficiently small (for which we expect the theory to be most accurate).

\section{Dark matter-wave solitons} \label{dark}

\subsection{Perturbation analysis }

We now consider repulsive BECs, which are characterized by a positive nonlinearity coefficient in Eq.~(\ref{eq:GP1}).  That is,  $g_{0}>0$, with $|\Delta g(z)|<g_0$ as in the previous case.  In the special case of a homogeneous nonlinearity, scaled so that its coefficient is $g_{0}=+1$, Eq.~(\ref{eq:GP1}) possesses a dark (black) matter-wave soliton solution of  the form
\begin{equation}
	\phi_{ds}(z,t)=\eta\tanh[\eta(z-z_{0})]\exp(i\mu t)\,,\label{eq:DS}
\end{equation}
where $\eta$ is the soliton's amplitude and inverse width, $z_{0}$ is its center position, and $\mu=\eta^{2}$ is its effective chemical potential.  In the case under consideration, the solution (\ref{eq:DS}) has nonvanishing boundary conditions at infinity, with the asymptotic limits $\phi_{ds}(z)\rightarrow \pm \eta$ as $z \rightarrow \pm \infty$.  Consequently, one can cast the problem of the piecewise-constant nonlinearity in the form of a perturbation, reminiscent of what was done for linear external potentials in Ref.~\cite{PelKev07}. The perturbed part of the Hamiltonian can be written formally as 
\begin{equation}
		H_{1}=\frac{1}{4} \int \frac{\Delta g(z)}{\Delta {g_0}}\left[
\eta^{4} -\phi^{4}(z)\right] dz\,. \label{phd}
\end{equation}
We remark that for the perturbed solutions (see the relevant figures in the numerical section below), it is not obvious that this integral converges. However, for the unperturbed solution (for which the solvability condition under the effect of perturbation is applied), it is certainly convergent.

Following Ref.~\cite{PelKev07}, one can show in a straightforward manner that for the solution to survive under the effect of perturbation (using the perturbation parameter $\epsilon \equiv \Delta g_0$), the function
\begin{equation}
	M'(s)=\frac{1}{2}\int_{-\infty}^{+\infty}\frac{d\Delta {g}}{dz}(z)\left[\eta^{4}-\phi_{0}^{4}(z-s)\right]dz
\label{mpd}
\end{equation}
must vanish.  As was the case for the bright matter-wave solitons we analyzed in the previous section, 
this condition implies that the center of the dark matter-wave soliton must be located at the center of either a region in which the nonlinearity coefficient is $g_0$ or at the center of one in which it is $g_1$.

The stability of the dark-soliton solution we constructed depends on the sign of the second derivative of $M(s)$ evaluated at the root $s_{0}$ of the first derivative (\ref{mpd}).  Based on the analysis of Ref.~\cite{PelKev07}, an instability will be present with one imaginary eigenfrequency pair if $\epsilon M''(s_{0})<0$ and with exactly one complex eigenfrequency quartet if $\epsilon M''(s_{0})>0$. 
As was the case for bright solitons (but with some important differences, which we discuss below), this instability is dictated by the translational eigenvalue, which needs to bifurcate from the origin as soon as the perturbation is present.  For $\epsilon M''(s_0)<0$, the relevant eigenfrequency pair moves along the 
imaginary axis, leading to an immediate instability associated with exponential growth of a perturbation along the associated eigendirection.  For $\epsilon M''(s_0)>0$, on the other hand, the eigenfrequency moves in principle along the real axis, which, however, is filled with continuous spectrum.  Consequently, as a result of the opposite signature (to these eigenfrequencies) of the translation mode, 
the eigenfrequency exits as a complex quartet, signalling the presence of an oscillatory instability. 
In fact, following Ref.~\cite{PelKev07}, we find that the relevant eigenfrequencies are determined by the characteristic equation 
\begin{align}
	\lambda^{2}+\frac{\epsilon}{4}M''(s_{0})\left(1-\frac{\lambda}{2}\right)=O(\epsilon^{2})\,, \label{dark1}
\end{align}
where the perturbation is given by Eq.~(\ref{gofz2}) and we recall that the eigenvalues $\lambda$ are related to the eigenfrequencies $\omega$ through $\lambda^2=-\omega^2$.

\begin{figure}[h]
\includegraphics[width=100mm,keepaspectratio]{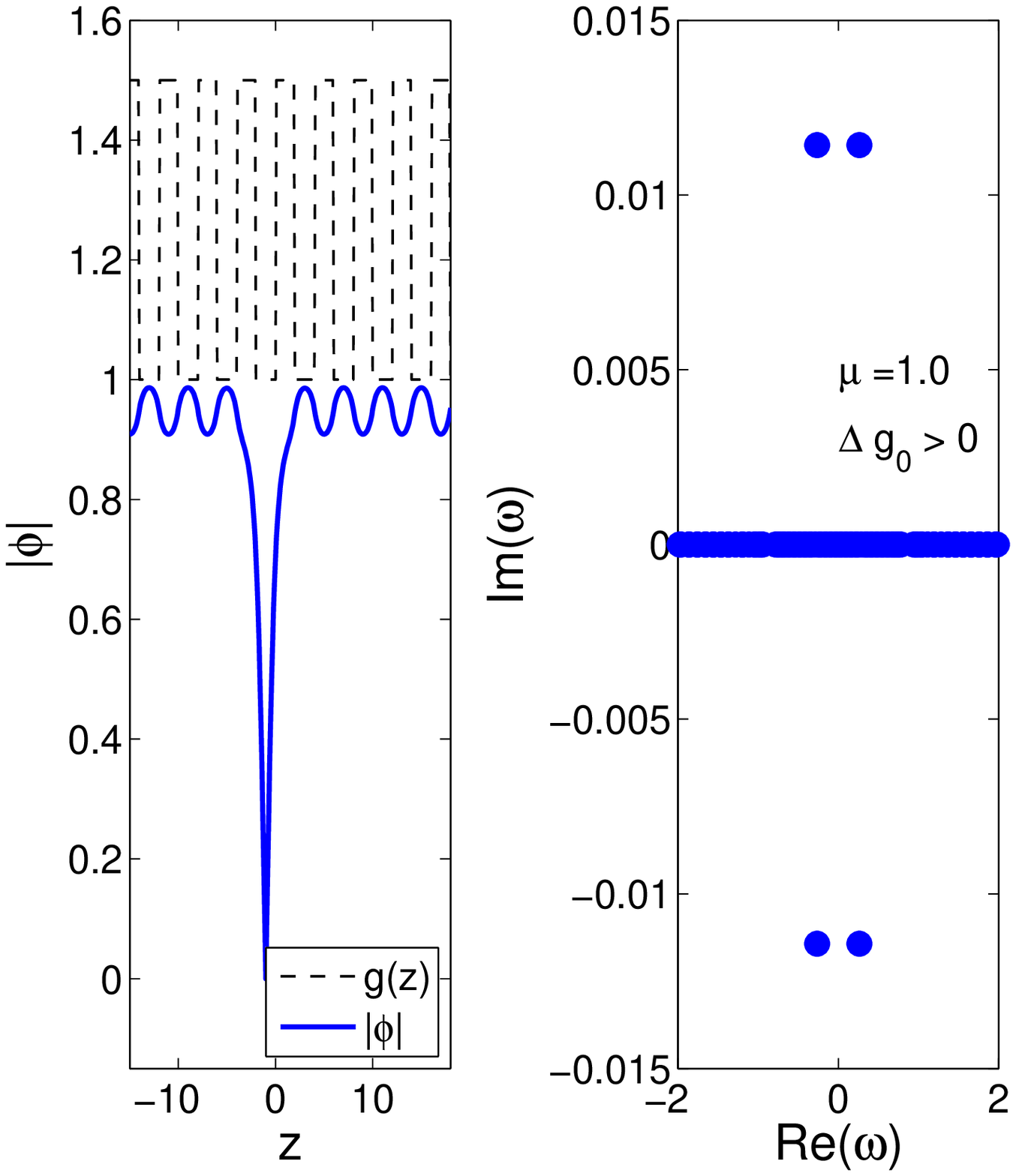}
\includegraphics[width=100mm,keepaspectratio]{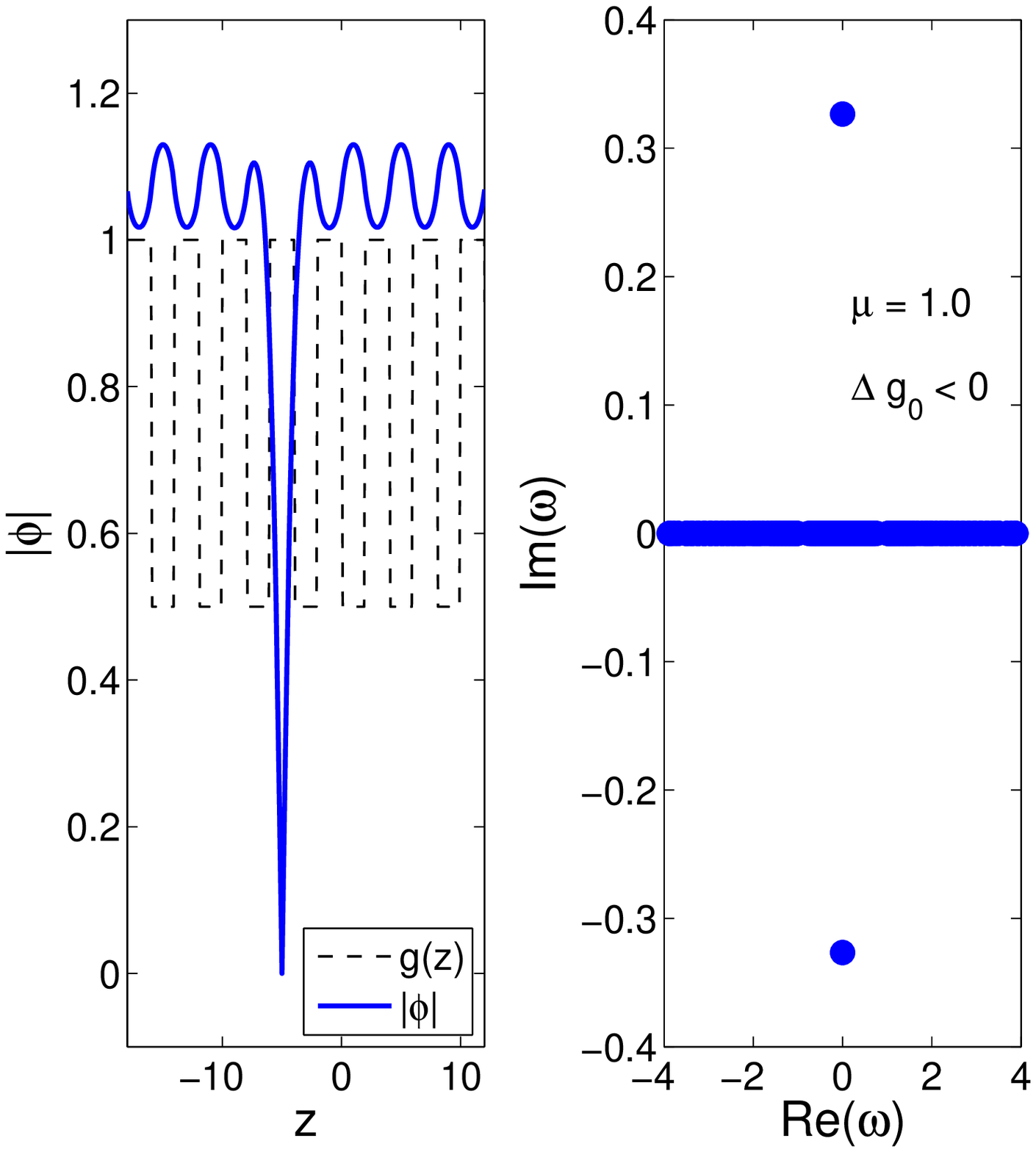}
\caption{(Color online) Numerical dark matter-wave soliton solutions of the NLS Eq.~(\ref{eq:GP1}).  
The top and bottom left panels show the wave profiles (solid curves) in the presence of a periodic, piecewise-constant coefficient in front of the nonlinearity (dashed curves).  In each case, the chemical potential is $\mu = 1$.  The soliton is centered at the $\min[g(z)]$ in the top panels and at $\max[g(z)]$ in the bottom ones.  The right panels show the eigenfrequency spectrum (similar to the right panels of Fig.~\ref{fig1}).  The top panel illustrates an oscillatory growth due to a complex quartet of eigenfrequencies, and the bottom shows an exponential growth due to an imaginary eigenfrequency pair. 
}
\label{fig3}
\end{figure}

Because the two roots of $M'(s)$ are $s_{01}=nL+L_{1}/2$ and $s_{02}=nL+L_{1} + L/2$,
one can evaluate $M''(s_{0})$ explicitly to obtain
\begin{align}
	M''(s_{01})=4\eta^{5}\sum_{p=-\infty}^{p=+\infty}\left\{ \tanh^{3}\left[\eta\left(pL+\frac{L_{1}}{2}\right)\right]
	-\tanh^{5}\left[\eta\left(pL+\frac{L_{1}}{2}\right)\right]\right\} \label{dark2}
\end{align}
and a similar expression for $M''(s_{02})$.  Combining Eqs.~(\ref{dark1}) and (\ref{dark2}) yields a
prediction for the location of the relevant translational eigenfrequency. To our knowledge, this is the first time that this has been done (albeit formally) for the case of a dark soliton in the presence of a spatially inhomogeneous nonlinearity. We now turn to numerical simulations in order to examine the accuracy of our analytical results.


\subsection{Numerical Results}

\begin{figure}[tbp]
\includegraphics[width=100mm,keepaspectratio]{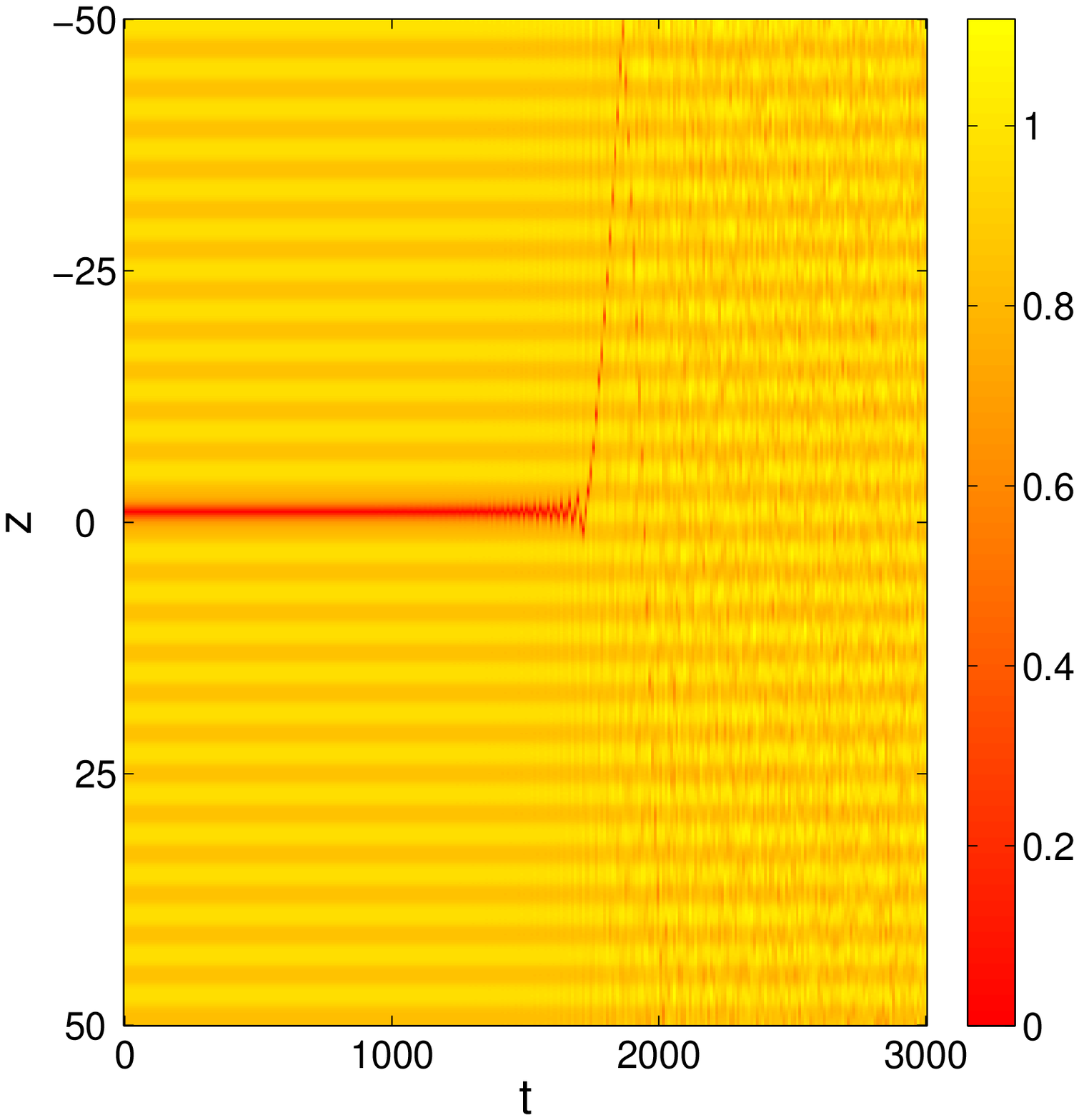}
\includegraphics[width=100mm,keepaspectratio]{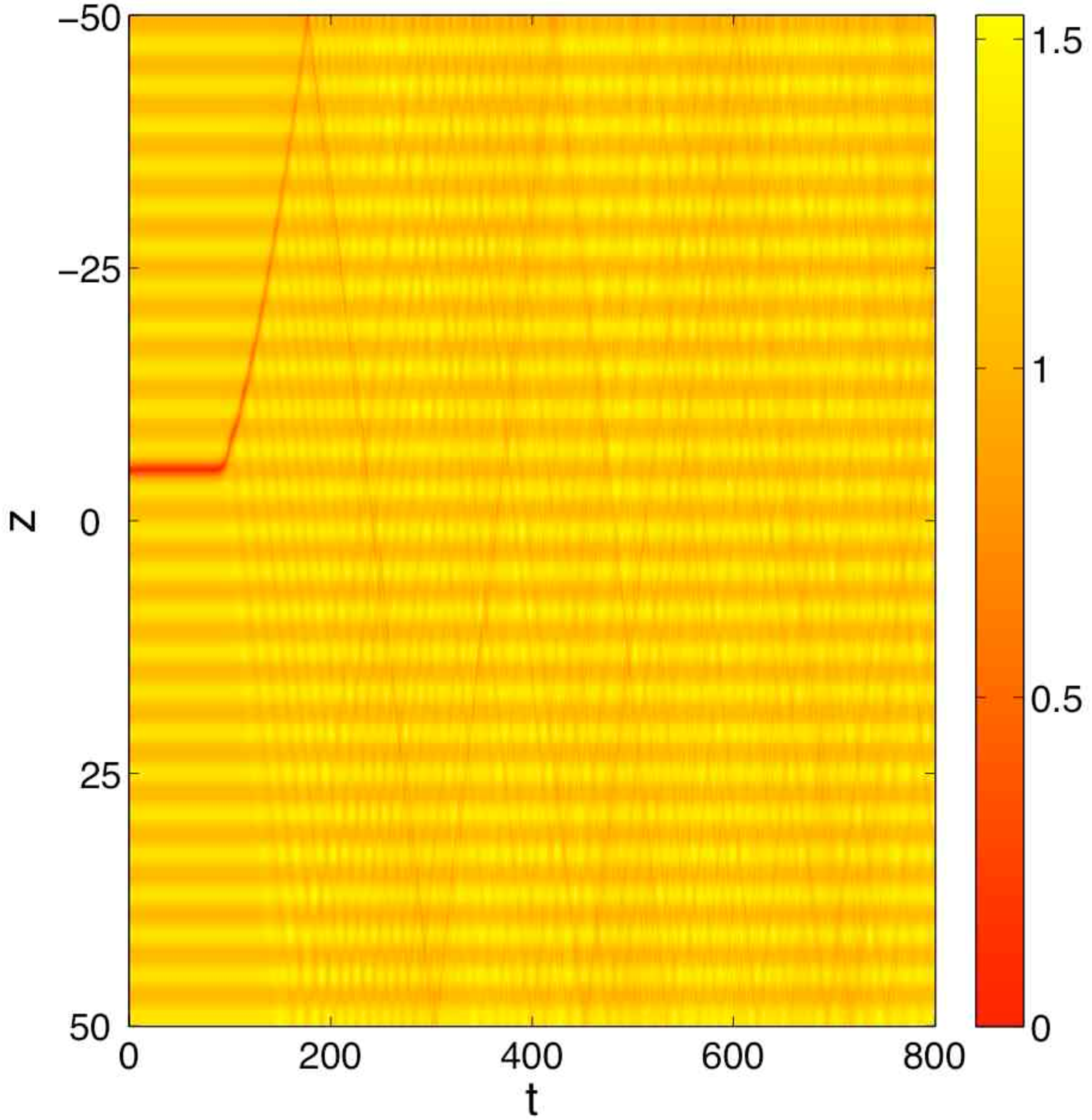}
\caption{(Color online) Time evolution of the dark soliton for the two configurations shown in the left panels of Fig.~\ref{fig3}.  The dark soliton shown in the top panel remains unchanged for a
longer time, showing evidence of an eigenfrequency with a smaller 
imaginary part. Furthermore, the former instability is oscillatory in its initial stages of development (indicating its association with a complex eigenfrequency quartet), whereas the latter
is not (indicating its connection to an imaginary eigenfrequency).}
\label{figDdyn}
\end{figure}

In Fig.~\ref{fig3}, we show the profile and spectral plane of the dark matter-wave soliton when it is centered at ${\min}[g(z)]$ (top panels) and ${\max}[g(z)]$ (bottom panels). The former case is associated with the scenario in which $\epsilon M''(s_0)>0$, and its corresponding spectral plane has an eigenvalue quartet (corresponding to an oscillatory instability). The 
latter case is characterized by $\epsilon M''(s_0)<0$, which results in
a real eigenvalue (imaginary eigenfrequency) 
pair in the spectral plane.

\begin{figure}
\includegraphics[width=100mm,keepaspectratio]{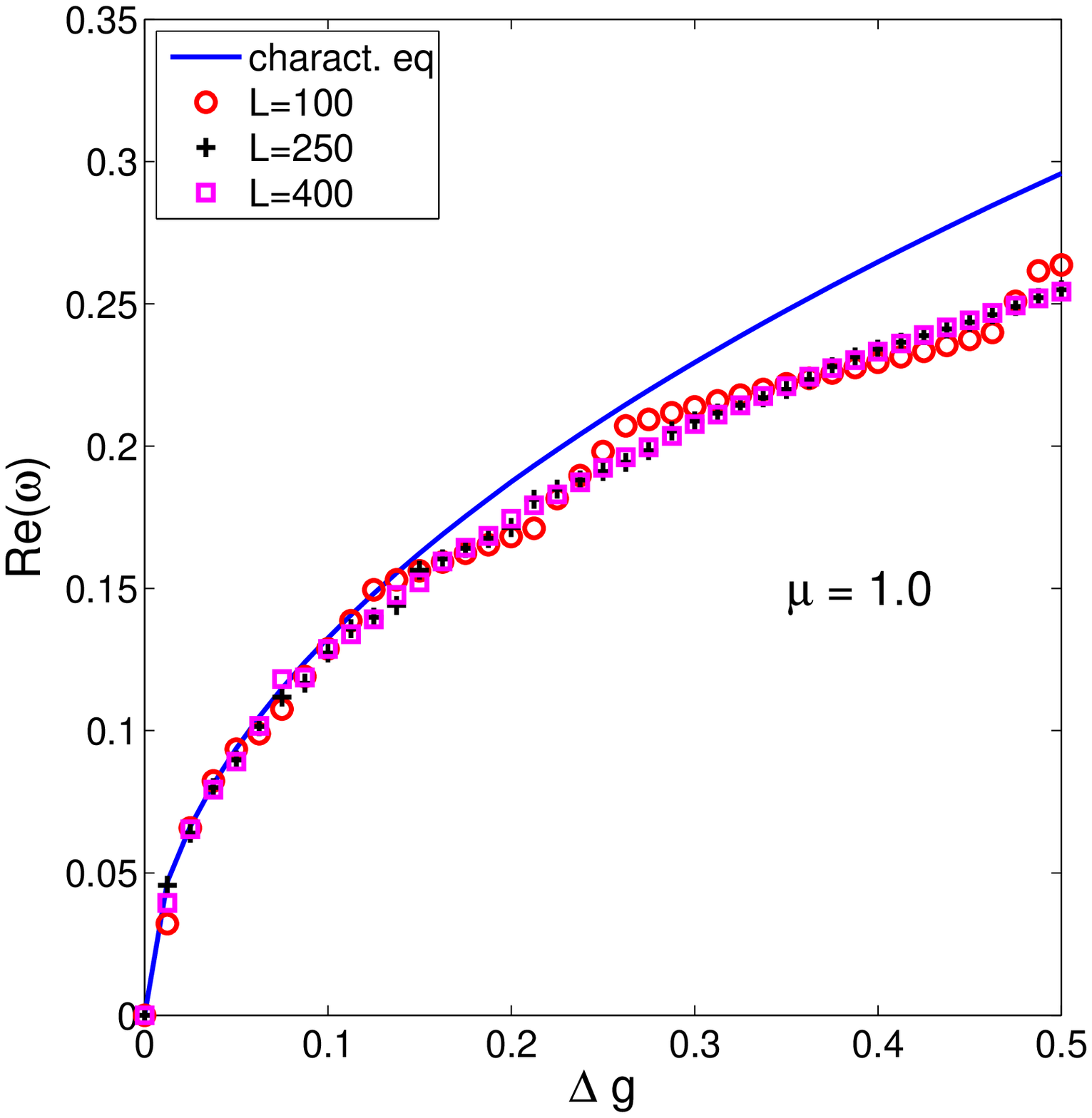}
\includegraphics[width=100mm,keepaspectratio]{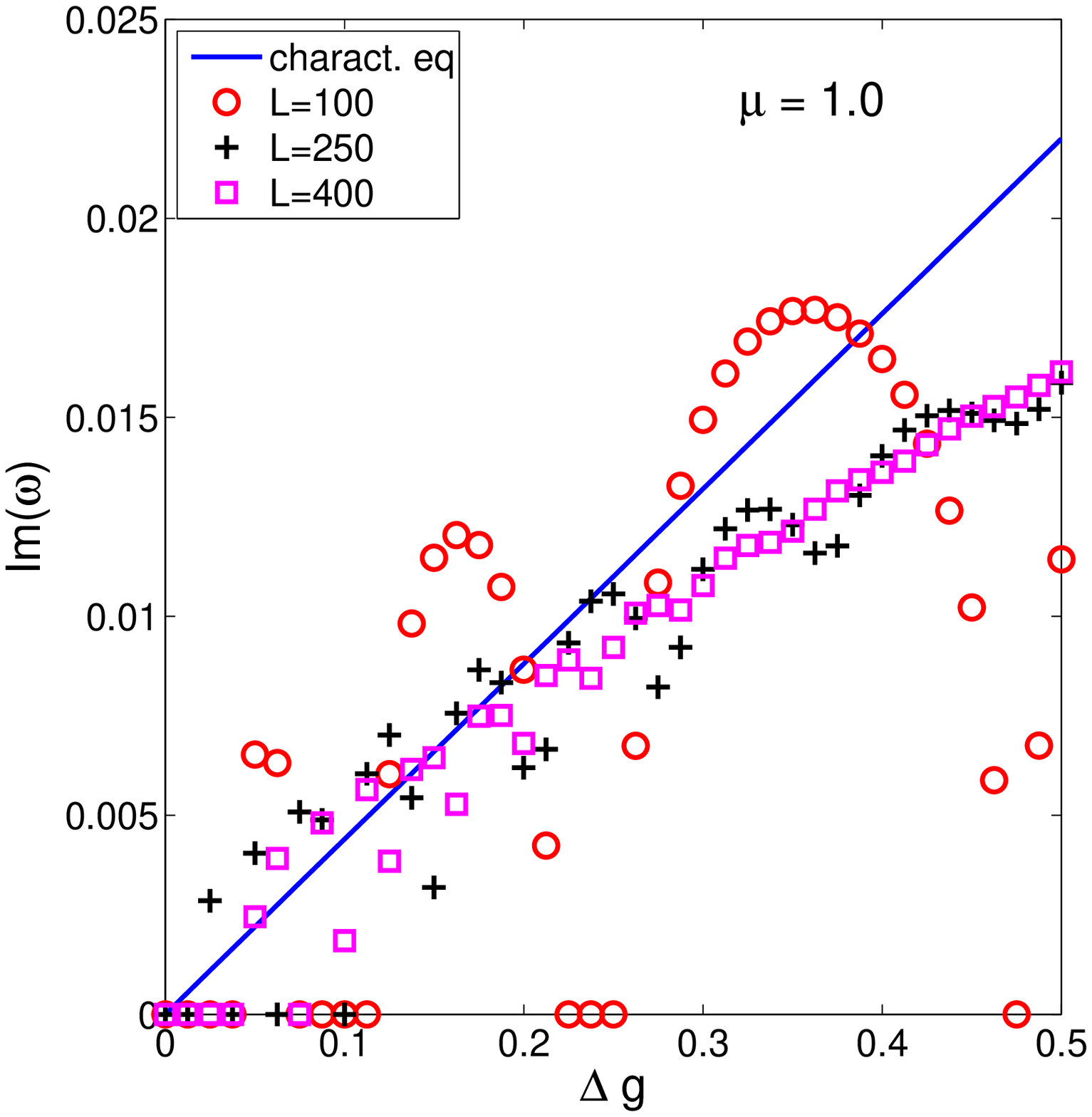}
\caption{(Color online) 
Real (top panel) and imaginary (bottom panel) parts of the eigenfrequency as a function of perturbation strength $\Delta {g_0}$ for a BEC with a chemical potential of $\mu=1.0$ which has a soliton centered at ${\min}[g(z)]$.  The points give the numerical results and the solid curve gives the theoretical prediction for the real and imaginary parts of the complex eigenfrequency quartet.  In both panels, we present the results of numerical simulations for three different domain sizes $L$ in order to illustrate the role  of the
finite size of the computational lattice, which is especially evident in the instability growth rate represented by {\rm Im}$(\omega)$.}
\label{fig4}
\end{figure}

\begin{figure}
\includegraphics[width=100mm,keepaspectratio]{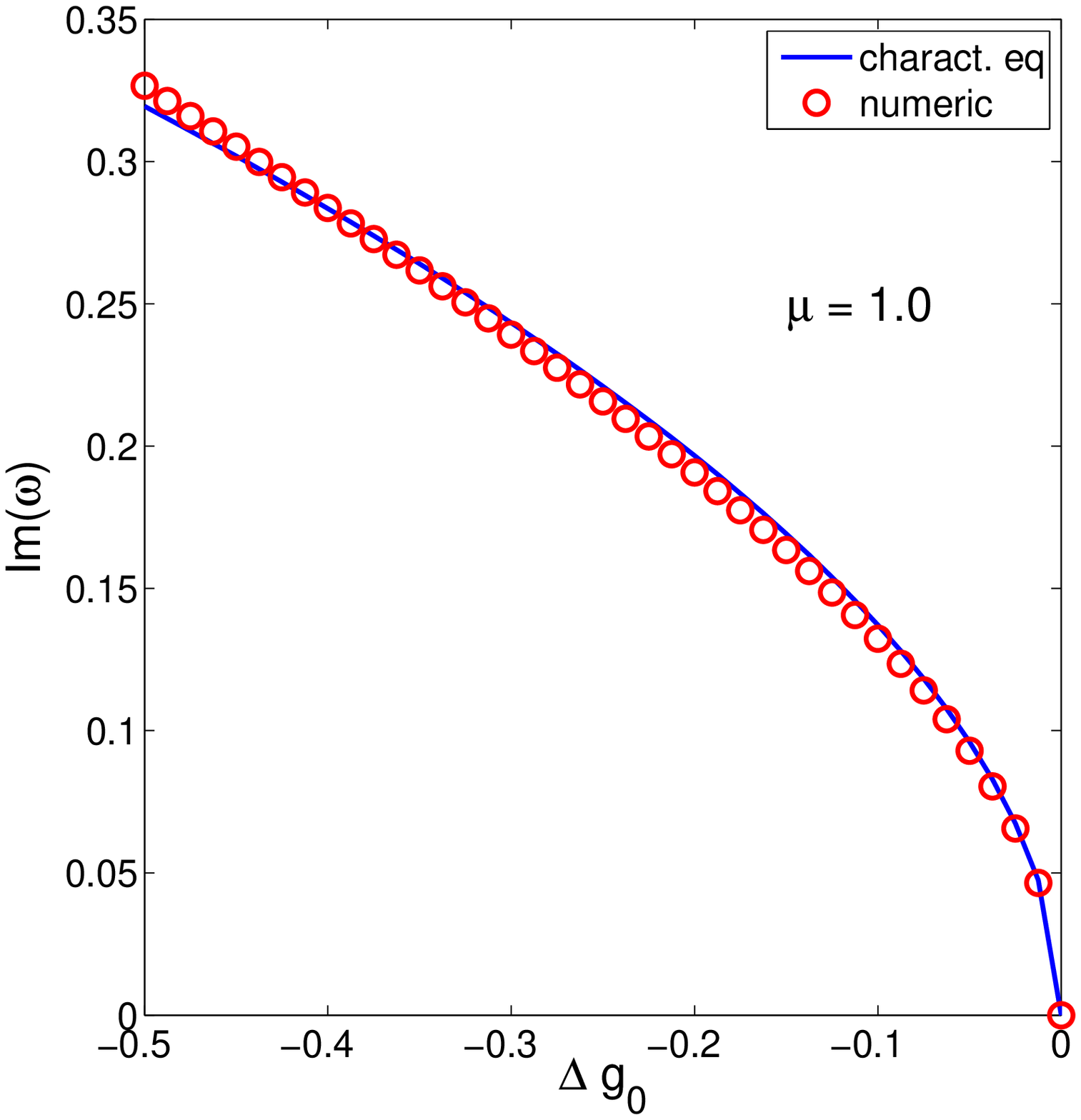}
\caption{(Color online) 
Imaginary part of the eigenfrequency as a function of the perturbation strength $\Delta {g_0}$ 
in for a BEC with chemical potential $\mu=1.0$ which has a soliton centered at ${\max}[g(z)]$. The
solid curve and the points correspond, respectively, to the analytical prediction and the numerical result for the imaginary eigenfrequency.}
\label{fig5}
\end{figure}

Figure~\ref{figDdyn} depicts the evolution of dark solitons with the two initial configurations shown in Fig.~\ref{fig3}.  In both cases, $\mu=-1$ and $|\Delta g_0|=0.5$.  In the top panel, the wave is centered
at ${\min}[g(z)]$, and in the bottom panel it is centered at ${\max}[g(z)]$.  Unlike their bright counterparts, dark solitons become mobile in both cases. This mobility develops earlier for the soliton in the bottom panel because its eigenfrequency has a much larger imaginary part of than that of the soliton in the top panel.  Furthermore, in the top configuration, the instability is initiated through an oscillatory phase that 
stems from its complex eigenfrequency quartet.  This feature is absent in the case of a wave centered at ${\max}[g(z)]$, as is expected when there is an imaginary eigenfrequency pair (which corresponds to a purely exponential instability). It is interesting to note that in both cases, as the soliton becomes mobile over the periodically nonlinear terrain, it radiates (a process that has been discussed elsewhere, although for a linear periodic potential; see, for example, Refs.~\cite{prouk} and references therein) and is eventually destroyed.

For the ${\min}[g(z)]$-centered soliton, we compare in the bottom panel of Fig.~\ref{fig4} the imaginary part of the eigenfrequency quartet (which corresponds to the real part of the eigenvalue quartet and yields the growth rate of the instability) to the theoretical predictions of Eqs.~(\ref{dark1})-(\ref{dark2}).  
Although finite-size effects similar to the ones reported in Ref.~\cite{PelKev07} (see the detailed discussion of Ref.~\cite{johaub2}) are present, the agreement between the theoretical prediction and the numerical result is good (and becomes better as the domain size increases).  We show three numerical cases and observe that the agreement with the theoretical prediction improves (as expected) as the density of the numerical phonon band of eigenvalues increases (for larger domain length $L$).  See Ref.~\cite{johaub2} for details on this finite size effect.  We also show the real part of the corresponding eigenfrequency for different domain lengths, for which the change in domain size does not play such
a significant role.  As shown in Fig.~\ref{fig5}, we observe a similar level of agreement (but without the finite size effects, because the relevant eigenfrequency is now imaginary) for a ${\max}[g(z)]$-centered dark soliton (for which $\epsilon M''(s_{0})<0$).

\section{Conclusions and Future Directions} \label{conc}

We have studied the existence and stability of bright and dark matter-wave solitons in Bose-Einstein condensates with spatially-dependent, periodic scattering lengths.  Our analysis was based on an analytically tractable model given by a Gross-Pitaevskii equation with a periodic, piecewise-constant, spatially inhomogeneous nonlinearity.  In particular, we used techniques from the theory of perturbed Hamiltonian dynamical systems in order to obtain conditions for the ``persistence'' of the matter-wave solitons and analyze their linear stability.  We thereby found that solitons must be centered in one of the constant regions of the piecewise-constant coefficient of the nonlinearity (which is directly proportional 
to the atomic scattering length).  We have shown that bright-soliton solutions are stable when localized in regions of maximal (absolute) nonlinearity and unstable when localized in regions of minimal  (absolute) nonlinearity. This is consistent with the findings of earlier works on the NLS equation with periodic potentials (see, for example, Refs.~\cite{key-2,key-4,ckrtj}).  In this situation, we also presented an approximate analytical technique (called ``stitching'') that allowed us to match the solution in a semi-analytical fashion at the interfaces where the nonlinearity coefficient changes in order to obtain an accurate profile of the soliton in the presence of the spatial inhomogeneity.  We applied a 
similar approach to dark matter-wave solitons (in that the translational eigenvalue is still responsible for the configurational stability) to show a rather different final result.  Specifically, we showed that dark solitons centered at ${\min}[g(z)]$ feature a complex eigenvalue quartet, whereas ones centered at ${\max}[g(z)]$, have a pair of unstable real eigenvalues.  Therefore, both cases are unstable, but the former instability is oscillatory and the latter is purely exponential (typically with a larger growth rate).  In all cases, we corroborated both the existence and the linear stability results with corresponding numerical computations, which were in good agreement with our analytical findings. 

It would be interesting to extend the present results to more complicated configurations. One interesting example along the lines of Ref.~\cite{ckrtj} might be to examine the competition of a piecewise-nonlinear and a piecewise-linear potential in one dimension.  Another possibility would be to investigate the stitching between spatially extended solutions given by elliptic functions.  Perhaps a more appealing example for a generalization of the present context would be to examine piecewise-constant nonlinearities in higher dimensions (with, for example, a piecewise-constant radial form or a piecewise-constant, square-lattice form) and determine the stability of vortex-like and similar structures in the
defocusing NLS.  In the focusing case, it would also be interesting to determine whether 
such a nonlinearity could stabilize solitary waves (which would be given piecewise 
generalizations of the Townes soliton) against collapse.

\section*{Acknowledgements}

A. S. R. acknowledges support from FCT through grants BSAB/695/2007 and PPCDT/FIS/56237/2004.
P. G. K. acknowledges support from NSF-DMS and CAREER and 
M. A. P. acknowledges support from the Gordon and Betty Moore Foundation through 
Caltech's Center for the Physics of Information (where he was a postdoc during much of this research). 
D. J. F. acknowledges support from the Special Research Account of the University of Athens. 
We thank Percy Deift, Dmitry Pelinovsky, and Bj\"{o}rn Sandstede for useful discussions.
Work at Los Alamos National Laboratory is supported by the US DoE.

\end{document}